# Towards Realistic Amorphous Topological Insulators

Marcio Costa,[†] Gabriel R. Schleder,[‡,†] Marco Buongiorno Nardelli,[¶] Caio Lewenkopf,[§] and Adalberto Fazzio[*,†,‡]

[†]*Brazilian Nanotechnology National Laboratory (LNNano), CNPEM, 13083-970 Campinas, Brazil*
[‡]*Center for Natural and Human Sciences, Federal University of ABC (UFABC), 09210-580, Santo André, São Paulo, Brazil*
[¶]*Department of Physics and Department of Chemistry, University of North Texas, Denton TX, USA*
[§]*Departamento de Física, Universidade Federal Fluminense, Niterói, Rio de Janeiro, Brazil*

E-mail: adalberto.fazzio@lnnano.cnpem.br

## Abstract

The topological properties of materials are, until now, associated with the features of their crystalline structure, although translational symmetry is not an explicit requirement of the topological phases. Recent studies of hopping models on random lattices have demonstrated that amorphous model systems show a non-trivial topology. Using *ab initio* calculations we show that two-dimensional amorphous materials can also display topological insulator properties. More specifically, we present a realistic state-of-the-art study of the electronic and transport properties of amorphous bismuthene systems, showing that these materials are topological insulators. These systems are characterized by the topological index $\mathbb{Z}_2 = 1$ and bulk-edge duality, and their linear conductance is quantized, $\mathcal{G} = 2e^2/h$, for Fermi energies within the topological gap. Our study opens the path to the experimental and theoretical investigation of amorphous topological insulator materials.

Advances in the synthesis and growth control at the nanoscale have allowed one to explore a variety of two-dimensional (2D) materials.[1] These systems show very unique electronic, transport and mechanical properties.[2,3] Of particular interest are 2D topological insulators,[4] that despite their experimental realization in HgTe/CdTe quantum wells[5] more than a decade ago, have been observed in very few systems so far.[6] Here, based on a realistic material modeling at the nanoscale, we propose a new path to investigate 2D topological insulators using amorphous materials.

The quantum Hall effect (QHE) initiated the era of topological states of matter.[7,8] The hallmark of QHE is the existence of a quantized Hall conductivity, first observed in a 2D electron gas under a strong magnetic field.[9] Inspired by the works of Thouless and Haldane,[10,11] Kane and Mele showed how a quantized conductance can arise for a class of Hamiltonian systems in the presence of spin-orbit coupling (SOC) without magnetic fields,[12] the so-called quantum spin Hall (QSH) effect. The QSH phase is characterized by the existence of metallic edge states protected against disorder by time-reversal symmetry ($\mathcal{T}$). These states carry a helical spin-polarization that counter-propagates along the edges (boundaries) of the system. In accordance with Kramers theorem, nonmagnetic impurities, which preserve $\mathcal{T}$, do not impair the quantized conductance of these metallic edge states.

Translational symmetry is an essential element in the standard definition of topological invariants that characterize topological insulator materials. If the translational symmetry is broken by disorder, one obtains systems that can no longer be classified by $k$-wave vectors. In non-crystalline solids or vitreous systems, due to the randomness of the potential, Bloch functions are strongly mixed, eventually leading to the appearance of localized states. A fundamental issue is whether a topological insulator protected by time-reversal is robust enough under a transformation to an amorphous state. The key role of translational symmetries in building the theory of topological insulators raises the question: to what extent is translational symmetry necessary for a topological state to retain its properties?

A non-trivial topological phase in an amorphous system was recently reported in a two-dimensional lattice of random set points simulating an amorphous Chern insulator.[13] This system was modeled by a two-orbital tight-binding Hamiltonian that can be regarded as a Bernevig–Hughes–Zhang (BHZ) model with an intra-orbital hopping.[14] Within the Chern insulator phase, ref [13] reported metallic edge states and a quantized conductance. Later, refs [15, 16] proposed the existence of a QSH phase in a 2D quasicrystal lattice (QL), which is also a non-periodic structure. Other studies suggesting the existence of non-trivial topology in amorphous systems include systems such as topological superconductors,[17] topological metals,[18] metamaterials,[19] and higher order topological insulators.[20] However, until now, topological phases in amorphous systems have only been reported for simple lattice models or non-material specific systems.

In this Letter, we propose the first realistic amorphous 2D material which features non-trivial topological properties. Our study is motivated by the recent synthesis report of the Bi monolayer (bismuthene) on a SiC substrate.[21] Bismuth is a heavy element with a large spin-orbit coupling[22] and its 2D allotropes are characterized by a rich phase diagram with competing trivial and topological phases.[23,24] The experimental synthesis of bismuthene on a SiC(0001) substrate via molecular-beam epitaxy (MBE) has been recently reported by Reis and coworkers.[21] These authors have used scanning tunneling microscopy (STM) and angle-resolved photoemission spectroscopy (ARPES) to probe the helical edge states of the sample, experimentally establishing the topological nature of bismuthene. These results nicely corroborate the theoretical prediction of a QSH insulating phase in bismuthene on a SiC(0001) substrate.[25] The latter uses first principles calculations and considers a pseudomorphic structure, where the Bi atoms acquire the substrate lattice parameter (5.35 Å), forming a honeycomb lattice.

The electronic structure of the amorphous materials studied in this Letter is computed in two steps. First, we perform Density Functional Theory (DFT)[26–28] calculations with plane waves basis sets implemented in the Quantum Espresso (QE)[29] software suite. The electronic exchange-correlation is described by the generalized gradient approximation (GGA) within the Perdew-Burke-Ernzerhof (PBE) functional.[30] Ionic cores are described using projector augmented wave (PAW) pseudopotentials.[31] This is the most computationally consuming part of the calculation and limits the system sizes we address here.

To study the topological and electronic transport properties we combine the flexibility of the plane wave basis with a novel methodology able to extract local effective Hamiltonians using pseudo atomic orbitals (PAOs), as implemented in the PAOFLOW code.[32–34] Our PAO Hamiltonians were constructed using a *spd* and *s* basis for Bi and H atoms, respectively. This allows us to efficiently address large system sizes. The resulting effective Hamiltonian reads

$$H_0 = \sum_{ij} \sum_{\mu\nu} \sum_{\sigma} t_{ij}^{\mu\nu} c_{i\mu\sigma}^{\dagger} c_{j\nu\sigma}, \quad (1)$$

where the operator $c_{i\mu\sigma}^{\dagger}$ ($c_{i\mu\sigma}$) creates (annihilates) an electron with spin projection $\sigma$ at the atomic site $i$ and orbital $\mu$. Here the Latin letters label atomic sites, while the Greek ones (except for $\sigma$) label the relevant atomic orbitals. The hopping matrix element $t_{ij}^{\mu\nu}$ is obtained directly from the DFT calculations, see supplemental material.[35] Note that due to amorphization, $t_{ij}^{\mu\nu}$ depends on the distance between the $i$ and $j$ sites and is direct obtained by PAOFLOW without fitting parameters. The atomic spin-orbit coupling (SOC) is introduced in the PAO Hamiltonian via an effective approximation, namely,

$$H_{\text{SOC}} = \sum_{i} \sum_{\mu\nu} \sum_{\sigma\sigma'} \xi_i^{\mu\nu} \langle i\mu\sigma | \boldsymbol{L} \cdot \boldsymbol{S} | i\nu\sigma' \rangle c_{i\mu\sigma}^{\dagger} c_{i\nu\sigma'}, \quad (2)$$

where $\boldsymbol{L}$ and $\boldsymbol{S}$ are the orbital and spin angular momentum operators, respectively. The SOC strength $\xi_i^{\mu\nu}$ is obtained from a fit to (relativistic) DFT calculations. This methodology has been successfully used in studies of other topological systems.[36,37]

To characterize the system topological properties, we calculate the $\mathbb{Z}_2$ invariant using the Wannier charge centers (WCCs) evolution method[38,39] implemented in the Z2Pack code.[40] We use PAO Hamiltonians as input to the Z2Pack code. The combination of WCCs and the PAO Hamiltonian is an extremely efficient computational method due to the PAOs compact subspace. With such an approach we were able to determine the topological nature of systems containing up to 400 atoms (4000 orbitals).

Finally, we also calculate the zero-temperature Landauer conductance $\mathcal{G}$ using the Caroli formula,[34,41–43] expressed in terms of the systems Green's function and the imaginary part of its embedding self-energy $\Sigma$, which describes the nanoflake coupling with the source and drain terminals. We calculate the embedding self-energy as in ref 34. The full system Green's functions, $G = (\epsilon - H_0 - H_{\text{SOC}} - \Sigma)^{-1}$, are obtained by direct inversion and expressed in the PAO basis.

Reference 25 has shown that the electronic structure and topological properties of the Bi/SiC(0001) and a hydrogen half-functionalized bismuthene (H-bismuthene) display essentially the same features. These materials are topological insulators ($\mathbb{Z}_2 = 1$) with an indirect band gap of 560 meV,[25] both presenting helical edge states with similar Fermi velocity. As a result of this similarity, we model the amorphous Bi/SiC(0001) using the H-bismuthene structure. We constructed a $10\times10$ hexagonal supercell with 400 atoms, 200 Bi and 200 H, to ensure a proper amorphous description.

We generate amorphous structures using the so-called bond-flipping procedure, recently used to study a honeycomb lattice with increasing concentrations of defects.[44] After each amorphization step we perform a full DFT structural optimization with a force threshold of $10^{-2}$ eV/Å, ensuring realistic amorphous structures. By employing this approach, the pristine bismuthene structure can be systematically driven towards an amorphous phase. The end result is a statistical distribution of polygons centered around hexagons (for details, see the supplemental material[35]), describing realistic amorphous systems.[45,46]

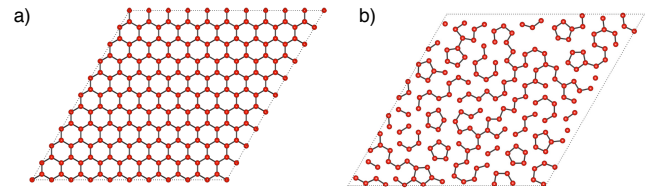

Figure 1: Atomic structure of the half H-passivated bismuthene: (a) pristine and (b) realization of an amorphous structure after DFT structural optimization.

The degree of amorphization can be qualitatively assessed by inspecting the lattice connectivity. Figure 1 shows a pristine bismuthene and a realization of an amorphous lattice obtained by the bond-flipping procedure, nicely illustrating how many bonds of the amorphous system are randomly disconnected in contrast to the pristine one (for details, see the supplemental material[35]). We illustrate two sites as disconnected when distances are longer than the pristine nearest-neighbor bond, i.e., 3.08 Å. The structural features and the amorphous character of the material can be further assessed by the pair distribution function (PDF) analysis[47] by studying the radial distribution function $g(r)$, defined by[48]

$$g(r) = \lim_{\Delta r \to 0} \frac{n(r)}{4\pi(N_{\text{pairs}}/V)r^2 \Delta r}, \qquad (3)$$

where $r$ is the distance between a pair of sites, $n(r)$ is the average number of site pairs found at a distance between $r$ and $r+\Delta r$, $V$ is the total volume of the system, and $N_{\text{pairs}}$ is the number of unique pairs of atoms. We compute $g(r)$ via the histogram method implemented in the VMD software.[49] The experimental relevance is discussed in the supplemental material.[35] Figure 2 shows the pair distribution function of the pristine and the amorphous structures corresponding to Figs. 1(a) and 1(b), respectively. The $g(r)$ of the pristine H-bismuthene, Fig. 2(a), presents sharp peaks regardless of the distance between the pair of sites. Conversely, in the amorphous phase, Fig. 2(b), only the peak centered at $d \approx 3.08$ Å, corresponding to the average distance between nearest-neighbor sites, is preserved. The vanishing peaks and the tendency of $g(r)$ to converge to a finite constant value for $r \gg d$ are manifestations of the absence of long-range order, expected for amorphous structures. In the insets, we show the fast Fourier transform (FFT) pattern of each structure, analog to a selected area electron diffraction (SAED) image. The presence of diffuse rings is characteristic of non-crystalline structures, in contrast to the sharp peaks of crystalline materials, such as in Fig. 2(a). In Fig. 2(b), a diffuse residual diffraction peak highlighted in red corresponds to a radial interatomic correlation of 4.7 Å, indicating a degree of local order expected by the statistical distribution of polygons centered on hexagons. The lack of additional peaks towards the image center, corresponding to longer distances, expresses the absence of long range order in the structure.

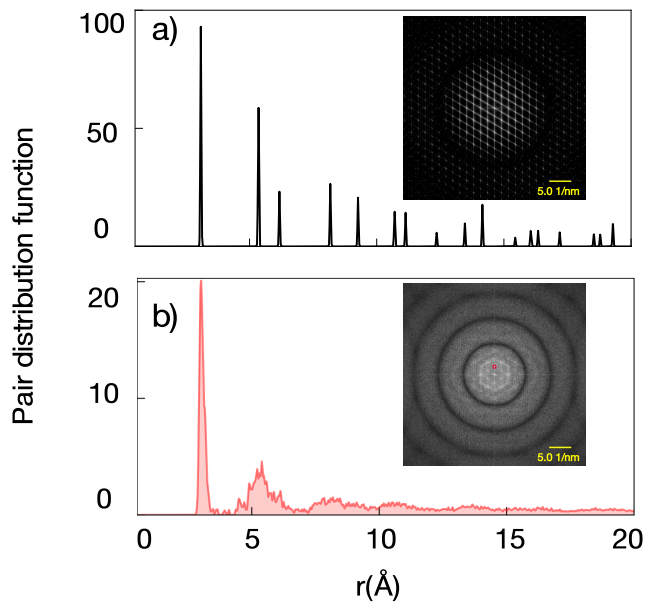

Figure 2: Pair distribution function (PDF) for half H-passivated bismuthene: (a) pristine and (b) amorphous. The inset of each panel shows the fast Fourier transform (FFT) pattern for each corresponding atomic structure in Fig. 1.

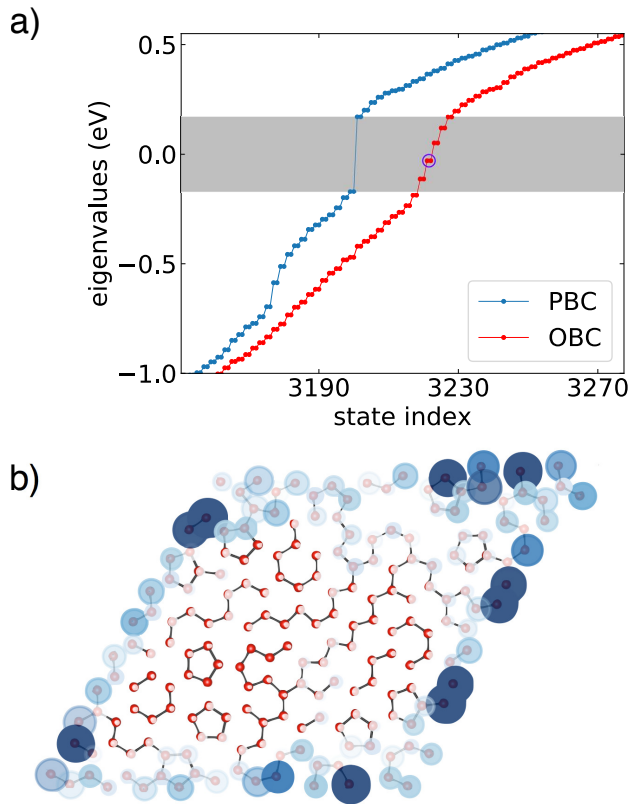

Figure 3: Amorphous H-bismuthene. (a) Energy levels (in eV) versus state index for periodic (PBC) and open (OBC) boundary conditions, corresponding to blue and red dots, respectively. The topological gap is indicated by the gray area. (b) Wave function spectral weight projected on the lattice sites of the nanoflake structure (OBC) for a representative mid-gap state, identified by the purple circle in panel (a).

To characterize the amorphous H-bismuthene topological properties we calculate the $\mathbb{Z}_2$ invariant using the results in the pseudo atomic orbitals (PAOs) representation. It has been shown that the topological phase transitions monitored by the $\mathbb{Z}_2$ and Chern number coincide with those observed with the spin Bott index and the Bott index, respectively.[16] As expected, we obtain for the pristine H-bismuthene $\mathbb{Z}_2 = 1$. Our calculations show that the amorphous H-bismuthene is also characterized by the invariant $\mathbb{Z}_2 = 1$, indicating its non-trivial nature.

Furthermore, the hallmark of a QSH insulator is the existence of topologically protected metallic (boundary) states once the system is interfaced with a trivial insulator, for instance vacuum. In Fig. 3(a) we show the energy levels for the amorphous H-bismuthene with periodic and open boundary conditions, which corresponds to a nanoflake geometry with 53.5 Å × 53.5 Å. In the periodic boundary condition case we find a 340 meV gap at $-0.17\,\text{eV} < E < 0.17$ eV. Once the periodicity is broken in all directions, we observe a number of states filling the energy gap of the periodic boundary case. All these mid-gap states that appear when considering open boundary conditions show a strong localization at the nanoflake edges, as expected for a topological material. Fig. 3(b) shows the site-projected wave function spectral weight of one representative mid-gap state. Due to the irregular edges, the wave function displays a significant variation along the edge, in contrast to edge states of pristine systems usually seen in crystalline QSH insulators. These results are in qualitative agreement with other works that study schematic models for amorphous topological insulators.[13,15–17,19]

Additionally, to probe the edge conductivity, we calculate the two-terminal electronic transport properties in a nanoribbon geometry, using the Landauer approach. In Fig. 4(a) we show the calculated conductance for an 800-atom nanoribbon having a width of 107 Å. The conductance features a quantized plateau of $\mathcal{G} = 2e^2/h$ in the energy gap region, $-0.12\,\text{eV} < E < 0.11$ eV, corroborating the picture of a QSH insulating phase. The site-projection of these mid-gap states is presented in Fig. 4(b), which reveals that the two conductive channels are located at the topological material edges.

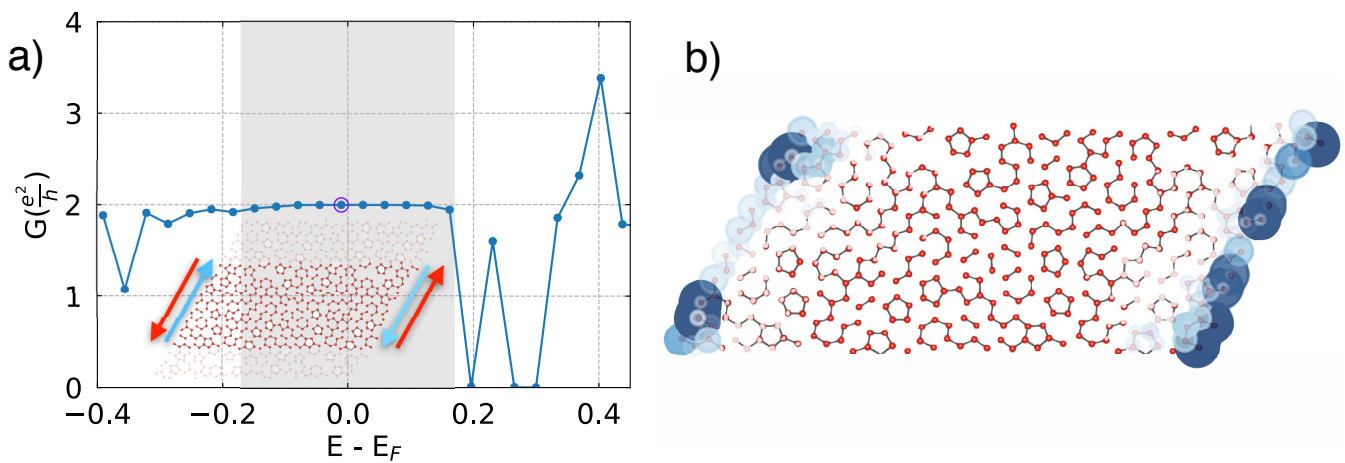

Figure 4: Amorphous H-bismuthene nanoribbon with a width of 107 Å. (a) Electronic two-terminal conductance, demonstrating the quantization plateau in the energy range of the topological gap (highlighted by the gray region). (b) Wave function spectral weight of one mid-gap state, highlighted in (a) by the purple circle, projected on the lattice sites, corroborating the localization of topological conductance channels at the edges of the system.

The lack of backscattering processes, manifested by a robust perfect conductance, the presence of edge states, and $\mathbb{Z}_2 = 1$ give solid evidence that, for $E_F$ within the topological gap, amorphous bismuthene is a topological insulator. On the other hand, for other values of $E_F$ we expect the system to behave as a trivial insulator. A quantitative characterization of the metal-insulator transition in terms of the localization length requires a multiscale analysis.[50,51] This is not possible within our approach, since the large computational cost, imposed restrictions on the system sizes we can deal with. To circumvent this limitation we use a simplified model to show that: (a) In the trivial phase, $\mathcal{G}$ is strongly suppressed with increasing nanoribbon length $L$, a fingerprint of the Anderson localization, while (b) in the topological regime $\mathcal{G}$ remains quantized regardless of the value of $L$. The details of these simulations can be found in the Supplemental Information.

Finally, we investigate the conditions under which the topological phase could be realized in other material systems. In Fig. 5 we evaluate the robustness of the topological phase as measured by the relative SOC parameter magnitude $\xi$. This value to a first approximation could represent other planar non-crystalline materials with smaller SOC, i.e., composed of lighter elements. The robustness is related to the minimum SOC necessary for a material to display a band inversion. In turn, the band inversion will occur if the material energy gap $E_g$ is sufficiently small so that SOC is able to invert it. This balance between the energy gap and SOC is a fundamental quantity to design novel amorphous topological insulators. For instance, smaller $\xi$ values can represent alloys of $Bi_{1-x}Sb_x$ that have been already experimentally synthesized[52,53] and can potentially show an amorphous topological phase. Other interesting possibilities involve also tuning either $E_g$ or $\xi$ by, for instance, application of strain or proximity effects, respectively. Such control allows one to build a topological phase diagram for a specific material, and the topological/trivial phase transition can be exploited for diverse applications purposes such as topological switches for electronic devices.

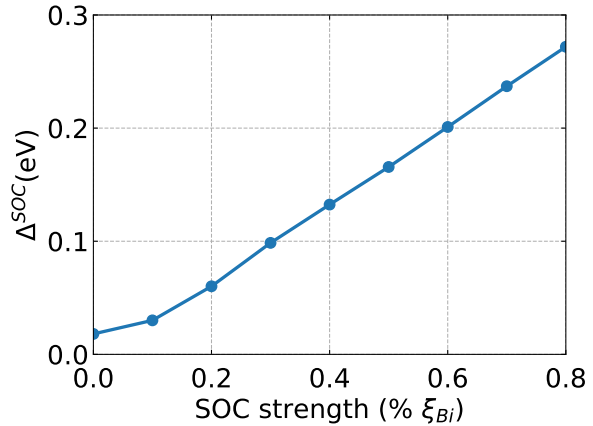

Figure 5: Topological phase diagram for the amorphous system as a function of the magnitude of the SOC parameter.

In this Letter we report the first realistic calculation of an amorphous material system with non-trivial topological properties. We generate an amorphous 2D lattice by applying an amorphization protocol starting from a hexagonal bismuthene structure, arriving at structures lacking long range order. We characterized the amorphous phase via the pair distribution function and the distribution of $n$-membered rings (pentagons, haxagons and heptagons). Our results are similar to the ones reported by recent experiments in other kind of 2D materials.[45,46] The electronic structure and transport calculations show existence of topological edge states both in amorphous bismuthene nanoflakes and nanoribbons, consistent with the bulk-edge duality. Moreover, the non-trivial phase is confirmed by the $\mathbb{Z}_2$ topological invariant and transport calculations showing the quantized conductance plateau in the topological gap. We expect this phase to be observed experimentally in the early stages of the bismuthene growth on SiC(0001), as the synthesis process uses an annealing treatment to increase the crystallinity of the obtained samples.[21] This strongly suggests the existence, before the annealing, of a substantially amorphous structure to be probed for topological states. Our results indicate the possibility of realizing topological phases in various non-crystalline systems and gives a strong motivation for the experimental verification of topological states in amorphous materials, opening the path for new potential applications.

**Acknowledgement** This work was partially supported by the Brazilian Institute of Science and Technology (INCT) in Carbon Nanomaterials and the Brazilian funding agencies FAPESP (Grants 16/14011-2, 17/18139-6, and 17/02317-2), CNPq (Grant 308801/2015-6), CAPES-PrInt (Grant 2561/2018), and FAPERJ (Grant E-26/202.882/2018). MC and MBN acknowledge Frank Cerasoli for useful discussions and technical support. MC, GS, and AF are thankful to Prof. Edson R. Leite for PDF analysis discussions. The authors acknowledge the National Laboratory for Scientific Computing (LNCC/MCTI, Brazil) for providing HPC resources of the SDumont supercomputer, and SAMPA/USP for the for providing HPC resources of the Josephson computer. MBN also acknowledges the High Performance Computing Center at the University of North Texas and the Texas Advanced Computing Center at the University of Texas, Austin, for computational resources.

# Supporting Information Available

Amorphization process, Amorphous structure ring size histogram, Tight binding and DFT band structures, pair distribution function details, and antimonene simulations.

# Graphical TOC Entry

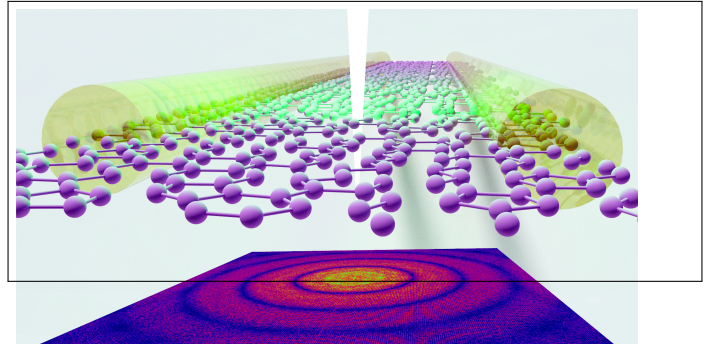